\documentclass[twocolumn,twoside,slac_two]{revtex4}
\usepackage{graphicx}
\usepackage{fancyhdr}
\providecommand {\etal }{et al. }%

\begin{document}

\title{Gamma-ray Clues to the Relativistic Jet Dichotomy}


\author{Eileen T. Meyer}
\email{meyer@rice.edu}
\affiliation{Department of Physics and Astronomy, Rice University,
    Houston, TX 77005}

\author{Giovanni Fossati}
\affiliation{Department of Physics and Astronomy, Rice University,
    Houston, TX 77005}

\author{Markos Georganopoulos}%\altaffilmark{2}}
\affiliation{Department of Physics, University of Maryland Baltimore County, Baltimore, MD 21250}
\affiliation{NASA Goddard Space Flight Center, Mail Code 663, Greenbelt, MD 20771, USA}
\author{Matthew L. Lister}%\altaffilmark{3}}
\affiliation{Department of Physics, Purdue University, West Lafayette, IN 47907}

\begin{abstract}
In examining a select sample of over 200 blazars of known jet kinetic
power ($L_\mathrm{kin}$) and well-characterized SEDs, we found
\citep{mey11} that Intermediate synchrotron-peaking (ISP) blazars may
have lower gamma-ray output than high synchrotron-peaking (HSP)
blazars of similar $L_\mathrm{kin}$, consistent with our hypothesis
that ISP blazars are less-beamed versions of HSP blazars, rather than
a distinct population. Further, by using the radio core dominance as a
measure of relative beaming, we find that gamma-ray luminosity depends
on beaming in a consistent way for blazars ranging over all jet
kinetic powers (10$^{42}$ $-$ 10$^{46}$ ergs s$^{-1}$). We re-examine
the gamma-ray properties of this core sample of blazars using the
1-year LAT catalog \citep{abd10_fermi_cat}.  We find that for weak
jets, the ratio of inverse Compton to synchrotron emission remains
constant with increased beaming, consistent with an SSC model for the
jet emission, while the most powerful jets show a strong increase in
Compton dominance with orientation, consistent with an external
Compton (EC) emission model.

\end{abstract}

\maketitle

\thispagestyle{fancy}


\vspace{-5pt}
\section{Introduction}
\vspace{-10pt}
Radio-loud active galactic nuclei (RL AGN) are believed to contain an
actively accreting super-massive black hole (10$^7$ - 10$^{10}$
$M_\odot$) which generates bi-polar jets of relativistic material
reaching up to Mpc in scale and with luminosities up to 10$^{49}$ ergs
s$^{-1}$, powerful enough to heat the intra-cluster medium and
potentially produce a feedback mechanism in galaxy formation
\cite{mac07}. When these relativistic jets are seen with the axis near
to the line of sight, they are called blazars (see \citep{urr95} for a
review). The jet radiation is strongly Doppler-boosted and dominates
over the emission from the host galaxy, the dusty torus (thought to
surround the nuclear black hole), the accretion disk, and, except at
very low frequencies, the isotropic emission from the giant lobes of
gas fed by the jet over time.  Thus blazars are naturally interesting
objects as the magnified view of the jet emission can allow us to
probe the structure and nature of these jets and ultimately to
understand the physical mechanisms behind their origin.

The broad-band, highly luminous and variable spectrum from blazars is
usually characterized as a `double-peaked' one, with one broad
emission component believed to be synchrotron radiation peaking from
sub-infrared energies to X-rays, and a second peak from inverse
Compton emission at $>$ MeV energies. This second high-energy peak in
blazars can be explained with both synchrotron-self Compton (SSC, in
which the particles in the relativistic jet up-scatter synchrotron
photons, \cite{mgc92,mar96}) and external Compton (EC) models, in the
latter case with photons from either the accretion disk, broad line
region (BLR), or molecular torus \cite{gm96,bla00,sik09}). Generally
low-power and lineless objects are believed to radiate by SSC at high
energies, while individually, many high-powered FSRQ have been more
satisfactorily fit with EC models \cite{ghi10,ver11}.  However a
consistent basis for which sources require EC has not been
demonstrated for any particular class of blazars.

The location of the GeV emission is also a matter of active debate, as
it comes from locations close to the central engine that remain
unresolved, even with very long baseline interferometry (VLBI). If the
emission region is located within the sub-pc scale BLR, the GeV
emission of blazars with strong lines is due to EC scattering of BLR
photons \cite{sik94}, which can lead to more efficient cooling and
thus lower peak frequencies, as seen in the blazar sequence
\citep{ghi98}. However, others agree that the high-energy emission may
be coming several pc downstream as suggested by some multiwavelength
observations of variability \citep{bot09,agu11}. The EC versus SSC
origin of the high-energy emission is an important question as it can
be used as a diagnostic for the location of the gamma-ray emission,
and the inferred structure of the jet.

%An additional complication is the increasing evidence of velocity
%gradients in the jet \cite{mey11,pin10}. 

The \emph{Fermi} gamma-ray telescope is giving us an unprecedented look at
the high-energy emission from blazars, with nearly 700 blazar
associations in the 1-year catalog (1-LAT)\cite{abd10_fermi_cat}. With
such a large sample, we can begin to put some order in the
phenomenology of blazars at high energies. In Section \ref{icenv} we
introduce the `Blazar Envelope' as seen by \emph{Fermi} and discuss
the importance of the jet kinetic power and orientation of the jet in
determining the gamma-ray luminosity.  In Section \ref{echp} we
discuss a recent finding suggesting that the high-energy emission of
high-power jets is dominated by EC rather than SSC. In Section
\ref{conc}, we summarize our findings.

\vspace{-5pt}
\section{The Inverse Compton Envelope as seen by \emph{Fermi}}
\vspace{-10pt}
\label{icenv}
The relativistic jets in RL AGN are generally confined to a small
opening angle and thus most blazars are seen within a small range of
orientation angles from the jet axis.  It is often assumed that bright
blazars in flux-limited samples are ``well-aligned'' and therefore the
exact viewing angle is of little importance (e.g., for modeling, often
the assumption $\theta$ = 1/$\Gamma$ is used). However, as surveys go
deeper, it is expected that more and more relatively misaligned
sources will be observed.

\begin{figure}
  \includegraphics[scale=0.5]{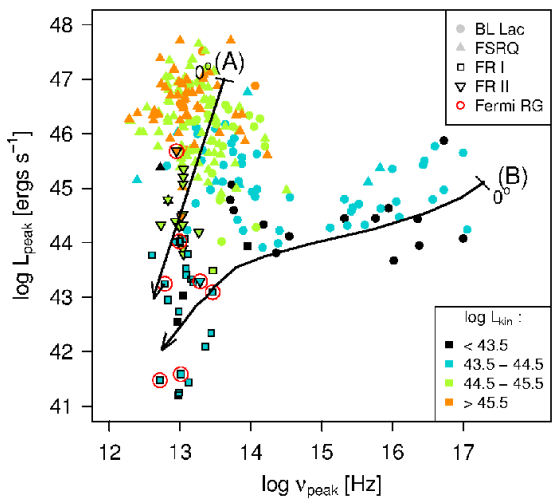}
  \caption{The synchrotron $L_{peak}$ - $\nu_{peak}$ plane, in
    which the blazar sequence first found an anti-correlation between
    bolometric luminosity (with $L_{peak}$) and the synchrotron
    peak frequency. Using only well-sampled SEDs reveals a possible
    alternative to the continuous sequence, in which two populations
    show dramatically different behaviors in this plane. Track (A) is
    the de-beaming trail in synchrotron peak for a standard single
    Lorentz-factor jet (`strong' type), track (B) shows the markedly
    horizontal movement typical of a decelerating jet model
    characterized by velocity gradients (`weak' type). \emph{Adapted from} \cite{mey11}.}
  \label{fig1}
\end{figure}
\begin{figure}
  \includegraphics[scale=0.5]{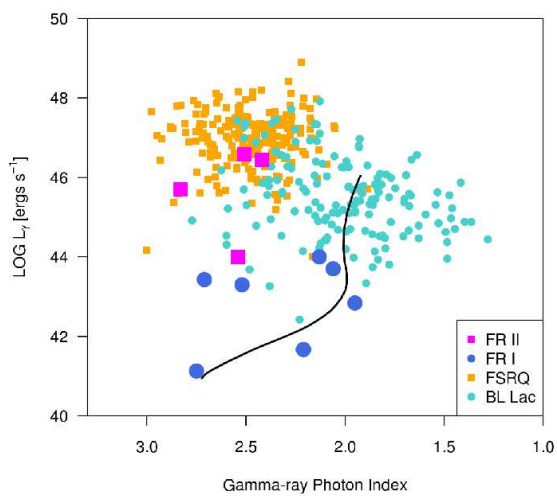}
  \caption{The Inverse Compton Envelope.  Figure is adapted from
    \cite{abd10_misaligned}.  The Fermi-detected AGN (large, dark blue
    circles and magenta squares) show that alignment plays a strong
    effect in the gamma-ray output.  Interestingly, while the FR II
    sources appear to drop directly below the powerful FSRQ sources,
    FR I radio galaxies appear to follow a more horizontal track,
    similar to our findings in the synchrotron envelope (Figure 1,
    left). The presence of a “forbidden zone” (empty region at upper
    right) suggests that there is a sequence in the IC plane, similar
    to that found in the synchrotron (Figure \ref{fig1}).}
  \label{fig2}
\end{figure}

Fossati et al. (1998) demonstrated an anti-correlation between the
synchrotron peak luminosity and the peak frequency, forming the now
canonical `blazar sequence' of (presumably) well-aligned sources
\cite{fos98}. Lower-luminosity sources appearing in the space below
the blazar sequence are expected due to some sources being less
aligned to the line of sight. Indeed, \cite{pad03} found that new
sources they identified modify the blazar sequence to an envelope,
with the area below the blazar sequence populated with
sources. Similar envelopes were found by Anton \& Browne and Nieppola
\etal \cite{ant05,nie06}.  In recent work, we filled the synchrotron
$L_\mathrm{peak}-\nu_\mathrm{peak}$ plane using a much larger sample
of jets over a wide range of orientations (from blazars to sources
misaligned enough to be seen as radio galaxies), with much-improved
SED sampling \cite{mey11}. This was the first work to consider both
the jet kinetic power ($L_\mathrm{kin}$, as measured from the
isotropic radio emission) and radio core dominance ($R$, a measure of
jet orientation), as important factors. Strikingly, we found that the
blazar sequence is actually broken into two populations in the
$L_\mathrm{peak} - \nu_\mathrm{peak}$ plane (Figure \ref{fig1}). The
`weak' jets consist entirely of sources with
$L_\mathrm{kin}<$10$^{44.5}$ ergs s$^{-1}$, and extend from
mis-aligned FR I radio galaxies at low $\nu_\mathrm{peak}$ up to
well-aligned (as measured by $R$) HSP BL Lacs. The `strong' jets
comprised a population of low $\nu_\mathrm{peak}$ jets which drop in
peak luminosity rapidly with decreasing $R$. The critical transition
between weak and strong jets appears to occur at an (dimensionless)
accretion rate of $\dot{m}_{cr} \sim 10^{-2}$ (see also
\cite{mg11_proc,ghi01}), as estimated from the ratio
$L_\mathrm{kin}$/$L_\mathrm{Edd}$\footnote{The Eddington luminosity was
  calculated using the average black hole mass estimate from the
  literature, as will be discussed in a forthcoming publication.},
matching the divide suggested by \cite{nar97}.

We note that the large number of blazars seen by \emph{Fermi}, as well
as the detection of several radio galaxies \citep{abd10_misaligned},
implies that we should have a large range in orientations in the
\emph{Fermi} samples. As see in Figure \ref{fig2}, an `envelope' is
indeed seen in the plane of gamma-ray luminosity ($>$ 100 MeV) versus
the LAT photon index (which approximately tracks the peak
frequency). There is clearly a `forbidden zone' at the upper right,
with the edge of the envelope presumably comprised of the most aligned
objects. The sources are divided into the typical higher-power and
broad-lined flat-spectrum radio quasars (FSRQ) and the lower-power,
lineless BL Lac objects (BLL). While this figure lacks the precise
information needed to find the two populations suggested by
\cite{mey11}, it is clear the the FR I radio galaxies (which are
unified with most BL Lacs) appear at much lower IC peak frequencies,
forcing a similar horizontal track from the low-power, aligned BLLs,
\emph{just as was found in the synchrotron plane}. It is interesting
to note that the FR II radio galaxies, which would be generally
unified with the FSRQ population, are both directly beneath them in
peak frequency, but are also not particularly mis-aligned, which
suggests that these sources might drop out of \emph{Fermi} detection
more quickly than the weak sources.

\begin{figure}
\includegraphics[width=65mm]{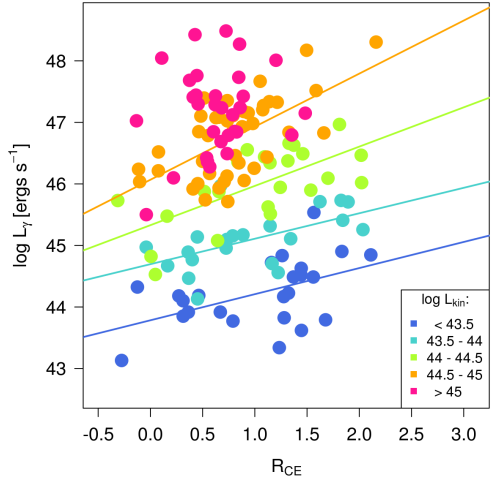}
\caption{Total gamma ray output versus radio core dominance for the
  130+ sources with good SED coverage detected by \emph{Fermi} in the
  1-LAT.  Jet power has a very strong effect on the gamma-ray
  luminosity, but when sources are sorted by their jet powers, the
  dependence on radio core dominance also becomes apparent.  The
  slopes shown are statistically significant at the 95\% level.}
\label{fig3}
\end{figure}

It is interesting to consider the effect of jet kinetic power as well
as orientation on the total LAT-band luminosity.  As shown in Figure
\ref{fig3}, as a total population the 1FGL blazars appear as a
scatter-plot in a plot of $L_\gamma$ versus $R$.  However, when the
jet kinetic power is shown (color scale), it becomes clear the
$L_\gamma$ increases with both radio core dominance (i.e., angle to
the line-of-sight) and, even more dramatically, with the jet kinetic
power.  We note that $L_\mathrm{kin}$ is calculated from the
low-frequency isotropic emission and is therefore a completely
independent measurement from $L_\gamma$, making it unlikely for this
trend to be the result of the selection effects of the 1-LAT sample.

\vspace{-5pt}
\section{Evidence for External Compton in High-Power Jets}
\vspace{-10pt}
\label{echp}

With a large sample of blazars spanning a range of orientations as
shown in the previous section, it is possible to look for signatures
of different high-energy emission mechanisms in a population of
blazars. The Doppler beaming factor, $\delta$, is a function of the
Lorentz factor $\Gamma$ and the orientation angle, and the effect of
$\delta$ on the observed (monochromatic) luminosity is:

\vspace{-15pt}
\begin{equation}
\mathrm{L} = \mathrm{L}_0\delta^{3+\alpha}
\label{eq2}
\end{equation}
\vspace{-10pt}

where $L_0$ is the rest-frame luminosity (at $\delta = 1$), and $\alpha$
is the energy spectral index at the frequency of interest.  The
exponent 3+$\alpha$ is the value assumed for a `moving blob' in the
jet. If the emission comes from a standing shock, the exponent would
be 2+$\alpha$ \cite{bla85}. 

In the case of emission by SSC, the IC peak has a beaming pattern
which is identical to the synchrotron (i.e., it follows Equation
\ref{eq2}). However, for EC models, the beaming at high energies goes
as $ \mathrm{L}_\mathrm{IC}$ =
$\mathrm{L}_\mathrm{0,IC}\delta^{4+2\alpha}$ \cite{der95,geo01}. The
larger exponent indicates that as a source with significant EC
emission is aligned, the IC peak should be more and more dominant over
the synchrotron peak (i.e., the ratio $L_\mathrm{IC}$/$L_\mathrm{sync}$
will increase with $R$. We can measure the Compton Dominance
$R_\mathrm{CD}$ = log($\mathrm{L}_\mathrm{IC}/\mathrm{L}_\mathrm{sync})
\propto $log ($\delta^{(3+\alpha)/(4+2\alpha)}$), and see that the dependence
on $\delta$ goes with an exponent of 2 when comparing the IC and
synchrotron peaks (where $\alpha$ = 1). At radio frequencies, where
the spectral index is confined to a fairly narrow range in values of
$\alpha_r = 0 - 0.5$, the beaming exponent will be $\sim$ 3 - 3.5.

\begin{figure}
\includegraphics[width=65mm]{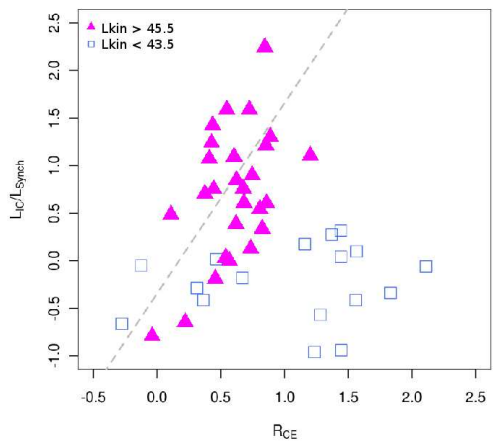}
\caption{We compare strong-jet and weak-jet sources in terms of their
  inverse Compton dominance over synchrotron emission. We find that
  for the strong jets, the inverse Compton dominance increases
  dramatically (up to a factor of 100) with increasing alignment, as
  measured by radio core dominance. Conversely, the weak jets show no
  such increase. The IC peak is measured by fitting the X-ray,
  gamma-ray, and TeV (where available) points with a parabola and
  taking the peak luminosity. (Dashed line shows slope=2 for
  reference).}
\label{fig4}
\end{figure}

In Figure \ref{fig4} we examine the relationship between
$R_\mathrm{CD}$ and the radio core dominance, $R$, for two sub-samples
of the 1-LAT blazars.  The pink triangles are high-jet-power sources
(log $\mathrm{L}_\mathrm{kin}$ $>$ 45.5), while the blue are low-power
(log $\mathrm{L}_\mathrm{kin}$ $<$ 43.5).  The flat distribution in the
latter group is consistent with the SSC models generally used to model
these types of sources. For the high-power group, however, we find a
significant trend of increasing Compton dominance with increasing
radio core dominance (i.e., alignment). \emph{This can be interpreted as
  a signature of an EC process for a population of jets characterized
  by high kinetic powers}.

However, from the above relations, the expected slope of the
correlation should be close to 2/3, up to a max of ~ 1 under
reasonable assumptions.  The slope in Figure \ref{fig4} is clearly
much higher, in fact consistent with a value of $\sim$ 2.  While a
more detailed investigation is in progress, we note that this high
slope might mean that the assumption that the Lorentz factor of the
plasma emitting in the gamma-rays is the same as that emitting at GHz
frequencies is incorrect.  If a roughly 2:1 ratio is assumed (i.e.,
$\Gamma_{IC}$ = 15 while $\Gamma_\mathrm{radio}$ = 8), the higher
slope can be explained.  This is consistent to some extent with the
difference in variability timescales for very high energies (minutes
to hours) versus radio (much longer), which indicates that the radio
emission is coming from a region which is slower and/or at larger
scales.

\vspace{-15pt}
\section{Conclusions}
\vspace{-10pt}
\label{conc}
We have shown that the \emph{Fermi} gamma-ray satellite sees
relativistic jets in RL AGN over a wide range of orientation angles,
including sources seen as radio galaxies.  There is a `forbidden zone'
of high-power, high-peak sources which remains empty, consistent with
a spectral sequence similar to that seen in the synchrotron peak
luminosity - peak frequency plane. Further, the much lower IC peak
frequencies found for the few detected FR I radio galaxies suggest
that the low-power class of FR I/BL Lacs debeam along more horizontal
tracks in the synchrotron and IC planes (Figures 1 and 2). We show
that the total gamma-ray band luminosity depends on both orientation
(measured through radio core dominance) and more heavily on the jet
kinetic power (as measured from low-frequency, isotropic radio
emission). Finally, we present the first collective evidence for EC
process in a sample of high-power jets (log $\mathrm{L}_\mathrm{kin}$
$>$ 45.5), showing that as alignment (radio core dominance) increases,
the Compton dominance increases dramatically.

\vspace{-10pt}
\begin{acknowledgments}
This research has made use of the SIMBAD database, operated at CDS,
Strasbourg, France (http://simbad.u-strasbg.fr/simbad/), and the
NASA/IPAC Extragalactic Database (NED) which is operated by the Jet
Propulsion Laboratory, California Institute of Technology, under
contract with the National Aeronautics and Space Administration
(http://nedwww.ipac.caltech.edu/). We utilized package \textit{np} in
the \textit{R} language and environment for statistical computing. GF
and EM acknowledge support from NASA grants NNG05GJ10G, NNX06AE92G,
and NNX09AR04G, as well as SAO grants GO3-4147X and G05-6115X. MG
acknowledges support from the NASA ATFP grant NNX08AG77G and NASA
FERMI grant NNH08ZDA001N. The MOJAVE project is supported under
National Science Foundation grant 0807860-AST.
\end{acknowledgments}

\bigskip % extra skip inserted

\end{document}